\documentclass[groupedaddress,twocolumn,pra,showpacs]{revtex4}
\usepackage{epsfig,amsfonts,amsmath,graphicx,subfigure}

\usepackage[letterpaper]{geometry}
\newcommand{\be}{\begin{equation}}
\newcommand{\ee}{\end{equation}}
\newcommand{\ba}{\begin{eqnarray}}
\newcommand{\ea}{\end{eqnarray}}

\def\bea{\begin{eqnarray}}
\def\eea{\end{eqnarray}}
\def\ben{\begin{eqnarray*}}
\def\een{\end{eqnarray*}}

\def\>{\rangle}
\def\<{\langle}

\newcommand{\fig}[1]{Fig.~\ref{fig:#1}}

\begin{document}

\title{Laser ablation loading of a
surface-electrode ion trap} 

\author{David R. Leibrandt, Robert J. Clark, Jaroslaw Labaziewicz, Paul Antohi, Waseem Bakr, Kenneth R. Brown, and Isaac L. Chuang}

\affiliation{Center for Ultracold Atoms, Research Laboratory of
Electronics, \& Department of Physics\\ Massachusetts Institute of
Technology, Cambridge, Massachusetts 02139, USA}

\date{\today}

\begin{abstract}

We demonstrate loading by laser ablation of $^{88}$Sr$^+$ ions into a
mm-scale surface-electrode ion trap. The laser used for ablation is a
pulsed, frequency-tripled Nd:YAG with pulse energies of 1-10 mJ and
durations of 3-5 ns. An additional laser is not required to photoionize the
ablated material. The efficiency and lifetime of several candidate materials
for the laser ablation target are characterized by measuring the trapped ion
fluorescence signal for a number of consecutive loads. Additionally, laser
ablation is used to load traps with a trap depth (40 meV) below where
electron impact ionization loading is typically successful ($\gtrsim$ 500
meV).

\end{abstract}


\maketitle


Trapped ions have been shown to be one of the most promising platforms for
large-scale quantum information processing (QIP). Recently, development has
begun on miniaturized and scalable ion traps \cite{Slusher:05, Monroe:06,
Hensinger:06, Seidelin:06, Britton:06}. While these efforts have met with
some success, current designs suffer from technical challenges such as a
relatively small trap depth and greater sensitivity to stray electric fields
compared with the traps used in previous QIP experiments. Both of these
problems can interfere with loading ions. For example, the loading of a
surface-electrode printed circuit board ion trap with electron impact
ionization presented in \cite{Brown:07} was hindered by stray charges until
buffer gas cooling was implemented and micromotion compensation performed. 
Photoionization has been used to load shallow ion traps \cite{Monroe:06,
Seidelin:06, Britton:06}, but it requires additional frequency stabilized
lasers which are not readily available for every ion species.  A new and
elegant method in which atoms are photoionized directly from a MOT has been
shown to efficiently load ions at a few mK \cite{Cetina:07}, but the laser
requirements are even more demanding than for standard photoionization
loading.

Laser ablation of a solid target has been used to load ion traps as early as
1981 \cite{Knight:81, Kwong:90}. Ablation is a process in which a
high-intensity laser strikes a surface, causing the rapid ejection of
material that includes neutral atoms, ions, molecules, and electrons
\cite{Phipps:book}. With other methods of ion loading, the neutral atoms are
ionized inside the trapping region.  This, however, is not the case with
ablation. It was shown in \cite{Hashimoto:06} that the electrons from the
ablation plume reach the ion trap first and short the trap electrodes for an
amount of time on the order of 10 $\mu$s, and the ions from the ablation
plume which are passing through the trapping region when the trap voltages
recover may be captured. A recent paper demonstrated an alternative way to
load ion traps with ablation which uses photoionization to ionize the
neutral atoms in the ablation plume as they pass through the trap region
\cite{Hendricks:07}.

Laser ablation loading is potentially advantageous for QIP for two reasons.
First, it is very fast: ions can be loaded with a single laser pulse in much
less than one second. And second, because the heat load is negligible small
ablation targets could be integrated with a multi-zone trap for localized
loading. Thus far, however, no work has been done to determine whether
ablation is a viable method for loading the miniaturized and scalable ion
trap designs proposed for large-scale QIP.

This paper examines ablation loading of a shallow, surface-electrode ion
trap similar to the designs proposed for large-scale QIP. We characterize
several candidate materials for the ablation target to determine which
materials are the most efficient and last the longest for loading
$^{88}$Sr$^+$, then proceed to find the minimum trap depth at which laser
ablation loading is possible in this trap.

\begin{figure}[h]
\begin{center}
\includegraphics[width=7cm]{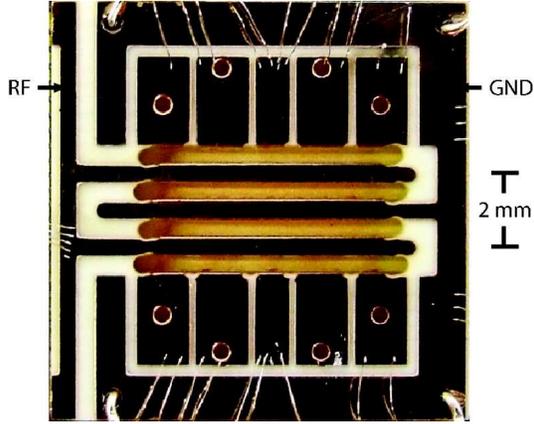}
\caption{ (color online) The
surface-electrode ion trap used for testing ablation loading. The RF
electrodes are spaced by 2 mm, leading to an ion height above the trap of
0.8 mm. The long center electrode is held at rf ground, but may have a dc
offset applied to it. The segmented electrodes on the sides carry dc
potentials for confinement along the long axis of the trap, as well as
elimination of stray electric fields.}
\label{fig:bastille}
\end{center}
\end{figure}
\begin{figure}[h]
\begin{center}
\includegraphics[width=7cm]{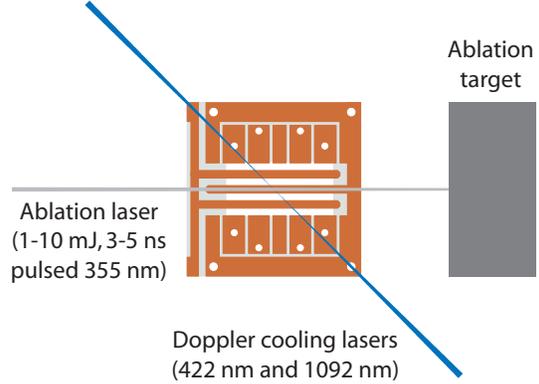}
\caption{ (color online) A diagram of the
setup showing the position and orientation of the ablation target relative
to the ion trap. The surface of the ablation target is approximately 25 mm
from the trap center and is orthogonal to the direction to the ion trap. Not
to scale.}
\label{fig:SMIT}
\end{center}
\end{figure}

The ion trap used for this work is a printed circuit board surface-electrode
Paul trap \cite{Chiaverini:05, Pearson:06, Brown:07, Cetina:07} shown in \fig{bastille}. The trap is
typically operated with 200-600 V rf amplitude at 8 MHz. The trap is mounted
in a ceramic pin grid array (CPGA) chip carrier, which is plugged into a
custom built ultra-high vacuum (UHV) compatible CPGA socket
\cite{Monroe:06}.  The socket is installed in a vacuum chamber evacuated to
~$2\times 10^{-9}$ torr. A schematic of the experimental setup is shown in
\fig{SMIT}.

We detect $^{88}$Sr$^+$ ions using laser-induced fluorescence on the 422 nm
5S$_{1/2}$ $\rightarrow$ 5P$_{1/2}$ transition, with a 1092 nm repumper beam
addressing the 4D$_{3/2}$ $\rightarrow$ 5P$_{1/2}$ transition to prevent
electron shelving in the metastable 4D$_{3/2}$ state. Fluorescence is
observed using either a photon counting photomultiplier tube (PMT) or an
electron-multiplying CCD camera.

The laser used for ablation is a pulsed, frequency-tripled Continuum
Minilite Nd:YAG laser at 355 nm. No additional photoionization lasers are
used. We load ions using a single laser pulse of energy 1-10 mJ and duration
3-5 ns. Ion numbers ranging from one to a few hundred are obtained with a
single pulse.



The efficiency of laser ablation loading is strongly dependent on the
ablation target material.  We studied several target materials by measuring
the trapped ion signal as a function of the number of ablation laser pulses
fired on a single spot of the target. Each ablation laser pulse knocks the
ions from the previous pulse out of the trap, so the trapped ion signal is
roughly proportional to the number of ions loaded by a single ablation
pulse. This measurement provides a benchmark of the loading efficiency and
the durability of the target.

\begin{figure}[h]
\begin{center}
\includegraphics[width=7cm]{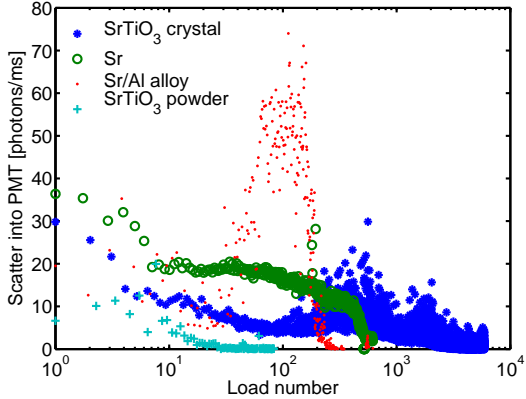}
\caption{ (color online) A plot of
the trapped ion signal as a function of the number of ablation pulses for
several different ablation targets. Each point represents the signal due
to a single ablation pulse of energy 8 mJ. For this experiment, the ablation
laser was focused to a spot size of 300 $\mu m$. For reference, a single ion
scatters roughly 2.5 photons/ms into the PMT in this setup.}
\label{fig:targets}
\end{center}
\end{figure}

The target materials studied here are Sr (99\% pure random pieces from
Sigma-Aldrich), Sr/Al alloy (10\% Sr, 90\% Al by mass from KB Alloys),
single crystal SrTiO$_3$ ($\left< 100 \right>$ crystal orientation from
Sigma-Aldrich), and SrTiO$_3$ powder in an epoxy resin (5 $\mu$m SrTiO$_3$
powder from Sigma-Aldrich mixed with Loctite 5 minute epoxy). In
\fig{targets} we plot experimental results for each of these targets. It is
clear that from a standpoint of durability and consistency that the
SrTiO$_3$ crystal is the best choice of target material for loading
$^{88}$Sr$^+$.  We are not concerned about the relatively lower efficiency
of SrTiO$_3$ because we are primarily interested in loading small numbers of
ions.

\begin{figure}[h]
\begin{center}
\includegraphics[width=7cm]{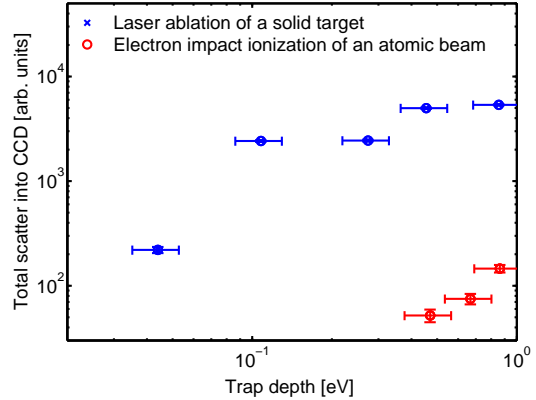}
\caption{ (color online) A plot of
the trapped ion signal as a function of the computed trap depth for both
ablation and electron impact ionization loading. An ablation pulse energy of
1.1 mJ was used with a spot size of 680 $\mu m$. Each point is the ion
signal obtained either from a single pulse of the ablation laser or from
loading using electron impact ionization until the ion signal stops
increasing.}
\label{fig:depth}
\end{center}
\end{figure}

We also measured the dependence of the trapped ion signal on the trap depth.
In this experiment, ions were loaded into the trap at a series of decreasing
rf voltages which correspond to decreasing trap depths. We calculate the
trap depth based on the time independent secular potential using a boundary
element electrostatics solver \cite{CPO, Pearson:06}, and verify that the
secular potential is accurate by checking that it gives secular frequencies
which match the experiment at each rf voltage. The trapped ion signal for
each trap depth is plotted in \fig{depth}. The ablation laser pulse energy
of 1.1 mJ and spot size of 680 $\mu$m were chosen to maximize the ion signal
at low trap depth. We found that the lowest trap depth at which we could
load using laser ablation is 40 meV. In contrast, the same experiment using
electron impact ionization of a thermal atomic beam loaded a minimum trap
depth of 470 meV.

The 40 meV trap depth loaded here with ablation is similar to the shallowest
trap depths loaded with photoionization of a thermal atomic beam
\cite{Monroe:06}.  Additional criteria to consider when selecting a loading
method for QIP include isotope selectivity and matter deposited onto the
trap electrodes. Matter deposited onto the trap electrodes is suspected to
increase the heating rate of the motional state of trapped ions
\cite{Turchette:99, Rowe:02}.  Photoionization loading is isotope selective
\cite{Kjaergaard:00, Tanaka:05} and deposits much less matter onto the trap
electrodes than electron impact ionization loading \cite{Gulde:01}.  The
isotope selectivity of ablation loading is similar to that of electron
impact ionization loading when loading the ions in the ablation plume as in
this work.  It is possible, however, to implement ablation loading in an
isotope selective manner by photoionizing the neutral atoms in the ablation
plume \cite{Hendricks:07}.  We have measured the ion heating rates in
cryogenic ion traps loaded with laser ablation and found them to be quite
low \cite{Labaziewicz:07}, which suggests that ablation does not deposit
much matter onto the trap electrodes.


In conclusion, we have used laser ablation of a solid target to load a
surface-electrode ion trap.  Several candidate materials for the ablation
target are characterized, and single crystal SrTiO$_3$ is found to give the
best performance for loading $^{88}$Sr$^+$. Laser ablation is demonstrated
to work for loading surface-electrode ion traps at trap depths as low as 40
meV.  When combined with the isotope selectivity and cleanliness
demonstrated elsewhere, these results suggest that laser ablation is a
viable loading method for large-scale ion trap QIP.

We acknowledge funding from Hewlett-Packard and the NSF.



\end{document}